\documentclass[preprint,12pt]{elsarticle}

\usepackage{amssymb}
\usepackage{amsmath}

\journal{Knowledge-Based Systems}

\usepackage{lineno}
\usepackage[utf8]{inputenc} 
\usepackage{amsmath, amssymb} 
\usepackage{algorithm}
\usepackage{algpseudocode}
\usepackage{graphicx} 
\usepackage{hyperref} 
\usepackage{booktabs} 
\usepackage{caption} 
\begin{document}

\begin{frontmatter}







\title{Prostate Capsule Segmentation from Micro-Ultrasound Images using Adaptive Focal Loss}

\author[a]{Kaniz Fatema}
\author[a]{Vaibhav Thakur}
\author[a]{Emad A. Mohammed\corref{cor1}}  

\ead{emohammed@wlu.ca}

\cortext[cor1]{Corresponding Author}

\address[a]{Department of Physics and Computer Science, Wilfrid Laurier University, 75 University Ave, Waterloo, ON, N2L 3C5, Canada}

\begin{abstract}
Micro-ultrasound (micro-US) is a promising imaging technique for cancer detection and computer-assisted visualization. This study investigates prostate capsule segmentation using deep learning techniques from micro-US images, addressing the challenges posed by the ambiguous boundaries of the prostate capsule. Existing methods often struggle in such cases, motivating the development of a tailored approach. This study introduces an adaptive focal loss function that dynamically emphasizes both hard and easy regions, taking into account their respective difficulty levels and annotation variability. The proposed methodology has two primary strategies: integrating a standard focal loss function as a baseline to design an adaptive focal loss function for proper prostate capsule segmentation. The focal loss baseline provides a robust foundation, incorporating class balancing and focusing on examples that are difficult to classify. The adaptive focal loss offers additional flexibility, addressing the fuzzy region of the prostate capsule and annotation variability by dilating the hard regions identified through discrepancies between expert and non-expert annotations. The proposed method dynamically adjusts the segmentation model's weights better to identify the fuzzy regions of the prostate capsule. The proposed adaptive focal loss function demonstrates superior performance, achieving a mean dice coefficient (DSC) of 0.940 and a mean Hausdorff distance (HD) of 1.949 mm in the testing dataset. These results highlight the effectiveness of integrating advanced loss functions and adaptive techniques into deep learning models. This enhances the accuracy of prostate capsule segmentation in micro-US images, offering the potential to improve clinical decision-making in prostate cancer diagnosis and treatment planning.

\end{abstract}






\end{frontmatter}


\section*{Introduction}
Prostate cancer is the most commonly diagnosed cancer among men. In 2018, there were approximately 164,690 new cases and 29,430 deaths from prostate cancer in the United States (US) \cite{Bray}. By 2022, the number of new cases had risen to about 268,490, making it the second leading cause of cancer-related deaths in men, with an estimated 34,500 deaths that year \cite{siegel2021cancer}. Some projections suggest that by 2030, there could be 1.7 million prostate cancer patients worldwide \cite{maddams2012projections}. Therefore, early detection of prostate cancer is crucial for improving survival rates and quality of life for patients. Early diagnosis of clinically significant prostate cancer can increase the 5-year survival rate from 31\% to nearly 100\% \cite{wolf2010american}.
In the US, approximately one million prostate biopsies are performed annually using transrectal ultrasound (TRUS) guidance \cite{loeb2011complications}. However, because TRUS imaging has limited accuracy, many clinical guidelines recommend combining targeted biopsies with blind systematic approaches to enhance diagnostic reliability \cite{mottet2021eau}.
Micro-ultrasound (micro-US) is a novel 29-MHz ultrasound technology that delivers 3-4 times higher resolution than traditional ultrasound. Several studies \cite{sountoulides2021micro, dias2022multiparametric} suggest that it offers comparable diagnostic accuracy to MRI at a lower cost. Furthermore, its real-time visualization removes the necessity for image fusion, which is responsible for up to 50\% of diagnostic errors in MRI-ultrasound fusion-guided biopsies, as observed by Williams and his colleagues. \cite{williams2022does}. Micro-US enhances the detection of clinically significant cancers and reduces the risk of disease progression at an earlier stage.  
Manual segmentation remains the most reliable method for accurately delineating the prostate capsule area. However, prostate segmentation on micro-US is cognitively challenging for three reasons: first, while urologists and radiologists typically use axial imaging, micro-US employs oblique planes. Second, the midline often has unclear boundaries between the prostate, bladder, and adjacent vasculature. Third, micro-US images can exhibit artifacts. Consequently, micro-US determines biopsy locations relative to the urethra (midline) rather than the capsule. Its varying shapes and sizes and the challenges of grayscale annotation require specialized training for surgeons. The difficulty distinguishing the prostate capsule from surrounding structures further complicates segmentation, making manual processes time-consuming and impractical for large populations. While micro-US improves accuracy, many urologists still prefer MRI for treatment planning due to its ability to generate 3D cancer maps. Therefore, precise and scalable prostate capsule segmentation is crucial for accurately locating prostate cancer concerning the gland.

The proposed method in this paper introduces an adaptive weighting scheme that dynamically adjusts the weights of the proposed segmentation model based on the difficulty of regions and the model's performance during training. This innovation enhances segmentation accuracy and robustness, making the model more accurate and effective for diverse patient datasets. The proposed method optimizes training processes and ensures adaptability to various imaging conditions by improving generalizability and efficiently allocating computational resources. Additionally, the scheme mitigates the impact of annotation variability, resulting in more reliable outcomes. These advancements directly benefit clinical applications, enabling precise diagnoses, better treatment planning, and improved patient outcomes while reducing healthcare costs.
The main contributions of this paper are summarized below.
\begin{enumerate}
        \item This paper introduces an adaptive focal loss function that dynamically adjusts its weighting scheme based on the patient micro-US images' specific challenges, such as prostate capsule fuzzy boundary and annotation variability. This dynamic approach enhances segmentation accuracy by focusing on the most difficult and clinically relevant regions.
        
         \item The proposed loss function outperforms existing loss functions, such as standard (i.e., PyTorch library built-in focal loss) focal loss and annotation-guided binary cross entropy (AG-BCE)\cite{jiang2024microsegnet}, especially in handling low-contrast, noisy, or anatomically variable regions. The method proposed in Jiang et al. \cite{jiang2024microsegnet} achieved higher Dice coefficients (DSC) and lower Hausdorff distance (HD) distances, confirming its accuracy and reliability.
         
         \item The proposed method in this paper demonstrates how an adaptive loss function can improve performance in complex medical imaging tasks, highlighting a shift from static to context-aware and data-responsive approaches.
    \end{enumerate}

\section*{Related Work}
\subsection*{Deep learning models for prostate segmentation}
Accurate segmentation of the prostate capsule is crucial for various clinical applications, such as diagnosing and planning treatment for prostate cancer. Technological advances have facilitated the development of reliable and precise MRI image segmentation methods. Likewise, notable advancements have been made in segmenting the prostate capsule using micro-US and TRUS imaging. Nonetheless, previous research struggled with attaining optimal segmentation accuracy and managing complex regions effectively.

 Chen et al.\cite{chen2021medical} introduced a modified 3D AlexNet for automatic prostate cancer classification using magnetic resonance images. The model was simplified to reduce parameters and training time and incorporated bottleneck structures, batch normalization, and global average pooling to enhance performance. Overfitting management techniques addressed the challenge of limited dataset sizes. The model achieved 92.1\% accuracy, with an AUC of 0.964, outperforming ResNet 50 and Inception-V4. However, limitations include sub-90\% segmentation accuracy, algorithm optimization challenges, and better multi-modal integration to reduce mismatches and improve results.

 Ghavami et al.\cite{ghavami2019automatic} conducted a quantitative comparison of six open-source segmentation algorithms tailored for prostate magnetic resonance imaging segmentation. The evaluated models included UNet, VNet, HighRes3dNet, HolisticNet, DenseVNet, and Adapted UNet. These algorithms were chosen for their relevance and reproducibility in clinical imaging tasks. The final segmentation was achieved by integrating the predictions from five models using a majority voting mechanism in each voxel. Performance was primarily evaluated using Dice scores, providing a reliable metric to compare segmentation accuracy across networks. The key limitations of the study included a restricted dataset size, possible overfitting due to hyperparameter optimization, and a primary focus on segmentation accuracy, which may have overlooked factors such as training and inference time. This research offers a foundational analysis of segmentation algorithms, emphasizing the importance of thorough evaluation and validation to ensure applicability in real-world clinical settings.

 Guo et al.\cite{guo2023prostate} proposed a modified UNet deep neural network for prostate capsule segmentation in Micro-Ultrasound images. Using a dataset of 2099 images (1296 annotated), the model achieved high Intersection over Union (IoU) scores: 95.05\% (training), 93.18\% (validation), and 85.14\% (test). A 10-fold cross-validation yielded an average IoU of 94.25\% and 9\% accuracy, demonstrating robust performance. This research highlights the potential of deep learning in enhancing surgical and diagnostic precision. However, the small sample size and potential annotation biases limit generalizability. 
 Jia et al.\cite{jia2017prostate} proposed a prostate segmentation method using a coarse-to-fine strategy combining probabilistic atlas-based segmentation (via DRAMMS) and Deep Convolutional Neural Networks (DCNNs) for fine segmentation. Validated on the PROMIS12 dataset using 5-fold cross-validation on 50 samples, the method achieved high accuracy: Dice Similarity Coefficient (DSC) 0.88 ± 0.04, Average Boundary Distance (ABD) 1.74 ± 0.42 mm, and HD (95\% HD) 5.00 ± 1.25 mm, outperforming traditional methods in DSC and 95\% HD. This approach highlights the effectiveness of combining atlas-based and deep-learning methods for prostate MR segmentation. Limitations include segmentation mismatches due to surrounding tissue complexities. 

 Karimi et al. \cite{karimi2019accurate} also proposed a CNN-based method for robust automatic segmentation of prostate CTV in TRUS images, overcoming challenges such as weak landmarks and artifacts. The technique employed adaptive sampling during training to focus on hard-to-segment images and leverages a CNN ensemble to estimate segmentation uncertainties. Uncertain segmentations are refined using prior shape information from a statistical shape model. This approach achieved a Dice score of 93.9 ± 3.5\% and an HD of 2.7 ± 2.3 mm, outperforming competing methods and significantly reducing segmentation errors. The study demonstrates that incorporating uncertainty estimation and prior shape knowledge can enhance CNN-based medical image segmentation, particularly in challenging cases.

 Peng et al.\cite{peng2022h} introduced the H-ProSeg method, a hybrid approach for robust prostate segmentation in transrectal ultrasound (TRUS) images. It uses a small number of radiologist-defined seed points to address challenges like ambiguous boundaries and anatomical variations. The method includes three subnetworks: a principal curve-based model, a differential evolution-based artificial neural network (ANN), and an ANN subnetwork for smooth contour description. Tested on 55 brachytherapy patients, H-ProSeg achieved a DSC of 95.8\%, a Jaccard coefficient of 94.3\%, and an accuracy (ACC) of 95.4\%, with minimal performance drop (up to 2.5\%) under noise. It outperformed current techniques, offering improved prostate cancer segmentation.

In the meantime,  Peng and his other colleagues \cite{peng2023automatic} proposed a two-stage method named Auto-ProSeg for prostate segmentation in TRUS images. They used an attention-based U-Net for initial contours and a refined principal curve-based approach with an evolutionary neural network. This hybrid method improved performance by 6\%, with more consistent results (standard deviation under 4.5\%) compared to over 5\% without refinement. Auto-ProSeg outperformed methods like U-Net++, which achieved a DSC of 89. 6\% with a standard deviation of 5. 8\%.

Several related works have significantly contributed to prostate capsule segmentation in micro-US images. However, these previous studies encountered limitations in segmentation accuracy or handling challenging regions. Meanwhile, Jiang et al.\cite{jiang2024microsegnet} proposed the MicroSegNet model for automated prostate segmentation in micro-ultrasound images. This model is based on the TransUNet architecture and incorporates a multi-scale deep supervision module and an annotation-guided binary cross-entropy loss (AG-BCE), which emphasizes difficult-to-segment regions during training. The performance metrics, including the DSC and HD, demonstrated that MicroSegNet significantly outperformed other models, achieving a DSC of 0.932 and an HD of 2.12 mm. However, this study faced limitations, including potential bias due to data collection from a single institution and the inclusion of only microscopic cases of extracapsular extension. Additionally, challenges in accurately segmenting the midline of the prostate were noted, which may improve with a more extensive training dataset.

\subsection*{Adaptive loss functions in image segmentation}

Several boundary-aware segmentation losses based on the distance transform map have been proposed. Zhu et al. \cite{zhu2019boundary} introduced a boundary-weighted segmentation loss (BWSL) function to make the segmentation network more sensitive to boundaries during segmentation. Kervadec et al.\cite{kervadec2019boundary} proposed a boundary loss formulated as a distance metric focusing on contours rather than regions. This approach addresses unbalanced segmentation challenges by emphasizing integrals over boundaries instead of regions. Karimi \& Salcudean \cite{karimi2019reducing} proposed three loss functions based on the distance transform of the segmentation boundary, aimed at minimizing the HD in medical image segmentation. Moreover, several researchers explored the adaptive loss function. For instance, Lin et al. \cite{lin2017focal} developed the focal loss to address the class imbalance in object detection tasks, which has been widely adopted in medical image segmentation for its effectiveness in challenging regions. However, its application to micro-ultrasound images may require further customization to handle the high resolution and calcification artifacts characteristic of these images, which can obscure boundary details and complicate segmentation. Kamnitsas et al.\cite{kamnitsas2017efficient} proposed an adaptive deep learning model for brain tumour segmentation, achieving significant improvements in accuracy and robustness by using an adaptive loss function that adjusts during training based on performance. While this approach demonstrated strong potential, its direct application to prostate segmentation in micro-US images may require adaptations to address the unique imaging characteristics and clinical requirements, such as handling high-frequency noise and artifacts specific to micro-US technology.

The primary objective of this paper is to improve the accuracy and robustness of prostate segmentation in micro-US images by addressing the limitations of existing methods. This is accomplished by developing multi-scale deep supervision, which captures intricate features across different scales to ensure accurate segmentation, even under challenging regions. The proposed adaptive focal loss function dynamically adjusts its focus based on image sample difficulty and annotation variability, emphasizing hard-to-segment regions without compromising overall performance.

The innovation lies in the adaptability of the loss function, which recalibrates, allowing the model to handle the variability and complexity inherent in micro-US imaging. This adaptability ensures strong performance under challenging regions while maintaining practicality for real-time clinical applications. Additionally, the method can be adapted to other medical imaging tasks with similar challenges, providing a solution that is both specialized and flexible.
This strategy bridges gaps in existing methods by integrating multi-scale deep supervision and the adaptive focal loss function. The proposed method improves segmentation accuracy and ensures the model's reliability and utility in clinical settings, paving the way for better medical outcomes.

\section*{Methods}
The proposed method in this section outlines a comprehensive approach for prostate capsule segmentation from micro-US images, leveraging adaptive focal loss function to address class imbalance and improve segmentation accuracy. This section offers a comprehensive summary of the details of the proposed method.

\subsection*{Algorithm: Adaptive focal loss for prostate capsule segmentation from micro-ultrasound images}

\begin{algorithm}[htbp]
\caption{Adaptive Focal Loss for Prostate Capsule Segmentation from Micro-Ultrasound Images}
\begin{algorithmic}[1]
\Require Micro-ultrasound image $I$
\Ensure Prostate capsule segmentation mask $S$

\State \textbf{Preprocessing:}
\Statex \hspace{1cm} a. Load input image $I$.
\Statex \hspace{1cm} b. Normalize pixel intensity values to $[0, 1]$.
\Statex \hspace{1cm} c. Apply augmentation (e.g., rotation, scaling, flipping).

\State \textbf{Model Initialization:}
\Statex \hspace{1cm} a. Initialize segmentation model with encoder-decoder architecture.
\Statex \hspace{1cm} b. Integrate adaptive focal loss for class imbalance.
\Statex \hspace{1cm} c. Use pre-trained weights or parameter initialization.

\State \textbf{Feature Extraction:}
\Statex \hspace{1cm} a. Pass $I$ through encoder to extract multiscale features $F$.
\Statex \hspace{1cm} b. Use dilated convolutions for enhanced spatial context.

\State \textbf{Adaptive Attention Mechanism:}
\Statex \hspace{1cm} a. Apply channel and spatial attention to refine features.
\Statex \hspace{1cm} b. Generate refined feature maps $F_{\text{refined}}$.

\State \textbf{Decoder and Segmentation:}
\Statex \hspace{1cm} a. Decode $F_{\text{refined}}$ using skip connections.
\Statex \hspace{1cm} b. Upsample maps to match $I$'s resolution.
\Statex \hspace{1cm} c. Generate raw segmentation mask $S_{\text{raw}}$.

\State \textbf{Loss Function and Optimization:}
\Statex \hspace{1cm} a. Define adaptive focal loss:
\[
L_{\text{adaptive}} = - \alpha_t (1 - p_t)^\gamma \log(p_t),
\]
\Statex \hspace{1cm} where $\alpha_t$ adjusts class weights, $\gamma$ focuses on hard pixels, and $p_t$ is predicted probability.
\Statex \hspace{1cm} b. Combine adaptive focal loss with Dice loss.
\Statex \hspace{1cm} c. Optimize with Adam and learning rate scheduler.

\State \textbf{Training:}
\Statex \hspace{1cm} a. Train the model on the labeled dataset.
\Statex \hspace{1cm} b. Monitor metrics (DSC, HD) on validation set.

\State \textbf{Inference:}
\Statex \hspace{1cm} a. Predict segmentation mask $S$ for $I$.
\Statex \hspace{1cm} b. Refine $S$ with postprocessing (e.g., morphological operations).
\Statex \hspace{1cm} c. Evaluate $S$ using metrics (DSC, HD).

\end{algorithmic}
\end{algorithm}

\subsection*{Proposed Micro-Ultrasound Segmentation Model}

\begin{figure}[ht]
    \centering
    \includegraphics[width=\textwidth]{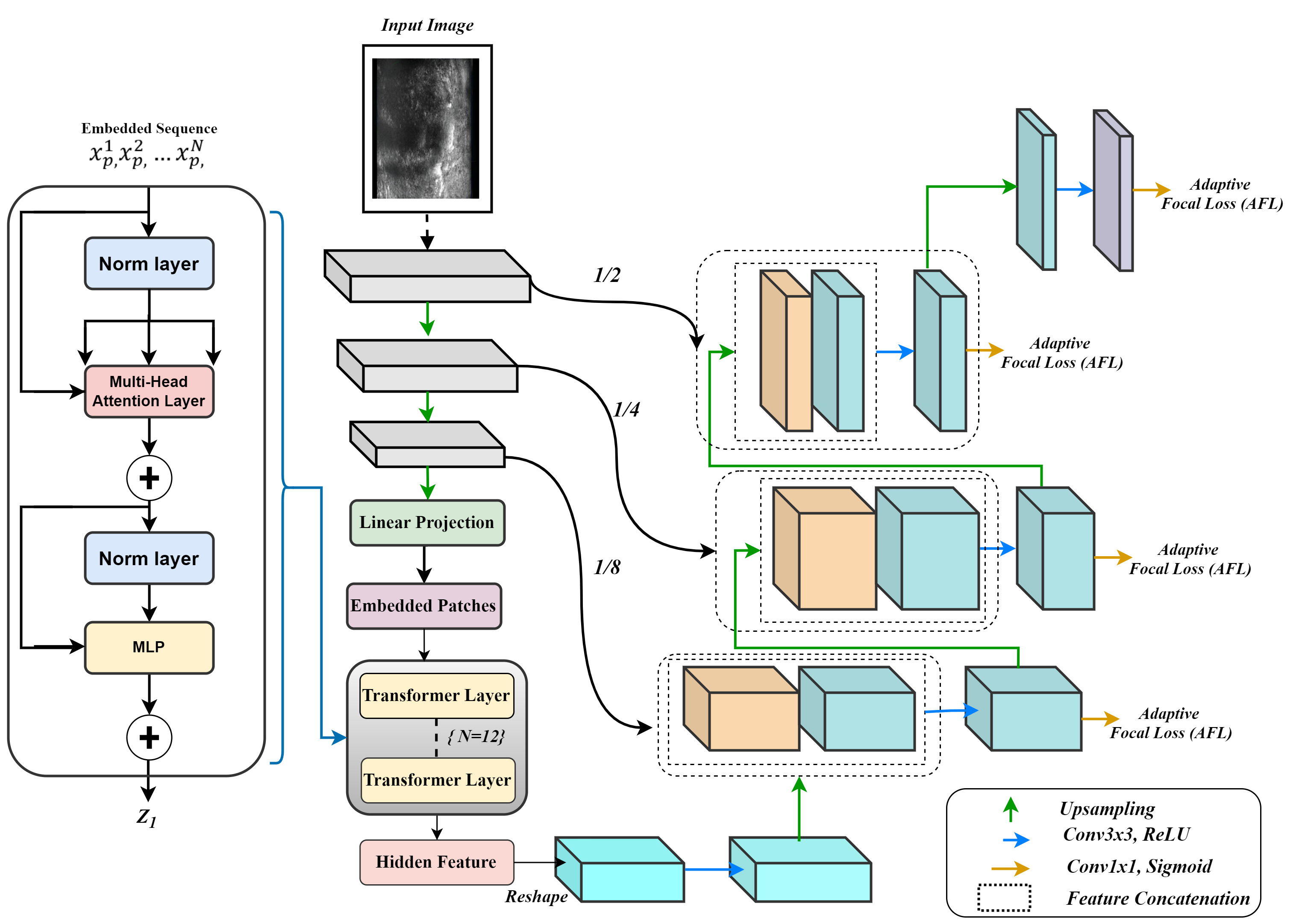}
    \caption{\textbf{Overview of the proposed segmentation model.} The model leverages the TransUNet architecture, with multi-scale deep supervision in intermediate layers for enhanced global context and the proposed adaptive focal loss in the output layer to improve accuracy, particularly in challenging regions.}
    \label{fig:Model}
\end{figure}

\noindent \textbf{Figure~\ref{fig:Model}} illustrates the architecture of the proposed segmentation model, which extends the TransUNet framework, a transformer-based approach specifically designed for medical image segmentation. Two critical modifications were implemented to improve the performance of the TransUNet model. First, a multi-scale deep supervision module was integrated to capture global contextual dependencies while preserving local information. Second, adaptive focal loss was incorporated into the training process to mitigate class imbalance by dynamically modulating the loss contribution of individual pixels. Collectively, these enhancements significantly improved the model's segmentation performance.

\subsubsection*{TransUNet Model Architecture}

The Vision Transformer (ViT) \cite{dosovitskiy2020image} is a transformer-based architecture originally designed for image classification. ViT processes images as tokenized patches and uses self-attention mechanisms to capture global contextual information effectively \cite{vaswani2017attention}. TransUNet is the first transformer-based model for medical image segmentation, integrating the strengths of ViT and UNet architectures. It combines global context modeling with high-resolution spatial details through a hybrid CNN-transformer design, enabling precise localization by merging self-attentive features with high-resolution CNN features. This combination achieves superior performance compared to traditional CNN-based segmentation models. The main components of TransUNet are described below.

\textbf{Convolutional stem.} A convolutional stem layer extracts low-level features from the input image while reducing spatial resolution.

\textbf{Image sequentialization.} Input features $\mathbf{x}$ from the convolutional stem are tokenized into a sequential representation of flattened 2D patches, denoted $\mathbf{x}_i^p \in \mathbb{R}^{P^2 \cdot C}$ for $i = 1, \dots, N$. Each patch has dimensions $P \times P$, and the total number of patches is $N = HW / P^2$, where $H$ and $W$ are the height and width of the input image. This process converts the input into a fixed-length sequence.

\textbf{Patch embedding.} The patch embedding layer divides the feature map into fixed-size patches $\mathbf{x}_p$ and projects them into a $D$-dimensional embedding space using a linear transformation. Positional embeddings are added to the patch embeddings to encode spatial information (equation 1):
\begin{equation}
\mathbf{z}^0 = [\mathbf{x}_1^p \mathbf{E}; \mathbf{x}_2^p \mathbf{E}; \dots; \mathbf{x}_N^p \mathbf{E}] + \mathbf{E}_{\text{pos}},
\tag{1}
\end{equation}

where $\mathbf{E} \in \mathbb{R}^{(P^2 \cdot C) \times D}$ is the patch embedding matrix, and $\mathbf{E}_{\text{pos}} \in \mathbb{R}^{N \times D}$ represents the positional embeddings.

\textbf{Transformer encoder.} The transformer encoder comprises \( L \) layers, each featuring multi-head self-attention (MHSA) and feed-forward networks (FFN), with residual connections and layer normalization. Patch embeddings are iteratively refined by attending to relevant patches. For the \(\ell\)-th layer, the output is (equation 2 \& 3):
\begin{align}
\mathbf{z}'_\ell &= \text{MHSA}(\text{LN}(\mathbf{z}_{\ell-1})) + \mathbf{z}_{\ell-1}, \tag{2}
\\
\mathbf{z}_\ell &= \text{FFN}(\text{LN}(\mathbf{z}'_\ell)) + \mathbf{z}'_\ell. \tag{3}
\end{align}
where \(\text{LN}(\cdot)\) is the layer normalization operator, and \(\mathbf{z}_L\) is the final encoded representation of the image.

\textbf{UNet decoder.} The UNet decoder consists of up-sampling blocks with convolutional layers, up-sampling layers, and skip connections to the convolutional stem layers. Patch embeddings are up-scaled to the original image resolution, and skip connections integrate high-resolution features from the stem with high-level features from the transformer encoder. The final output is a segmentation map, labelling each pixel in the input image.

\subsubsection*{Multi-scale deep supervision}

The TransUNet model is a baseline for reducing discrepancies between predicted segmentation maps and ground truth. However, it may face challenges related to input image shape, size, and appearance variations. Multi-scale deep supervision is implemented to address these issues, integrating information from multiple scales during the segmentation process \cite{xie2015holistically}. Each network pathway emphasizes features at a specific scale, enabling the model to combine local and global information.

To enhance performance, $1 \times 1$ convolutional layers with Sigmoid activation are applied to intermediate features at three scales ($1/2$, $1/4$, and $1/8$). This generates prostate segmentation maps at varying resolutions, where lower resolutions capture global context and higher resolutions focus on local details. For accurate predictions at all scales, ground truth segmentation maps are downsampled, and the differences between each prediction and its corresponding downsampled ground truth are incorporated into the loss function. By supervising predictions across multiple scales, the model effectively integrates features at different levels, capturing both high-level and detailed contextual information.

\subsubsection*{Proposed Adaptive\_Focal\_Loss}

The AG-BCE loss function \cite{jiang2024microsegnet} assigns weights to "hard" and "easy" regions in the data based on annotations. However, it uniformly treats these regions, disregarding inherent sample difficulty and annotation variability. This fixed weighting may constrain the model's learning potential, while adaptive weights can enhance learning, particularly in complex datasets with high variability or uncertain regions. An adaptive AG-BCE loss function aims to improve the model's focus on challenging regions.

Incorporating sample difficulty into the loss function is crucial, as different samples pose varying challenges. Samples near decision boundaries or with high prediction uncertainty are inherently more challenging. Integrating difficulty into the loss function ensures better attention to these cases, improving overall performance. This approach aligns with research by Lin and his teammates \cite{lin2017focal}, which introduced Focal Loss to emphasize hard-to-classify examples by scaling contributions based on model confidence.

Annotation variability presents another challenge, often reflecting disagreements or uncertainty among annotators. Regions with significant variability require distinct learning strategies compared to consistently annotated areas. Considering this variability helps the model focus on uncertain regions, consistent with findings by Kendall et al. \cite{kendall2017uncertainties}, who explored model uncertainty in loss functions for tasks like semantic segmentation and object detection.

To make the loss function adaptive, both sample difficulty and annotation variability are integrated into a dynamic "gamma" factor, which adjusts the influence of hard and easy regions. This dynamic adaptation enables the AG-BCE loss to effectively address the challenges of each sample, resulting in robust learning outcomes. Similar approaches, such as Dynamic Loss Weighting (DWG) \cite{chen2018gradnorm}, emphasize adapting loss functions based on task or sample significance.

This approach addresses inherent data challenges and enhances performance in complex and uncertain environments by synthesizing these concepts into an adaptive focal loss. The adaptive focal loss dynamically adjusts the gamma factor based on real-time assessments of sample difficulty and annotation variability. This method surpasses traditional approaches like AG-BCE, particularly in datasets with high complexity and uncertainty, aligning with recent advancements in adaptive loss functions. The pseudocode  for the adaptive\_focal\_loss is given below:

\paragraph {Pseudocode: Adaptive\_Focal\_Loss function}
\begin{enumerate}
    \item \textbf{Calculate the number of pixels:}
    \[
    N_{\text{pixels}} = \text{height} \times \text{width} \times \text{channels}
    \]

    \item \textbf{Identify hard regions:}
    \[
    \text{hard}_{\text{np}} = y_{\text{true}_{\text{np}}} \oplus y_{\text{std}_{\text{np}}}
    \]
    where \(\oplus\) denotes the bitwise XOR operation.

    \item \textbf{Perform dilation:}
    \[
    \text{hard}_{\text{dilated}} = \text{dilate}(\text{hard}_{\text{np}}, \text{kernel})
    \]
    with a kernel of size \( ks \).

    \item \textbf{Define easy regions:}
    \[
    \text{easy} = 1 - \text{hard}_{\text{dilated}}
    \]

    \item \textbf{Apply Sigmoid activation:}
    \[
    y_{\text{pred\_sigmoid}} = \text{torch.sigmoid}(y_{\text{pred}})
    \]

    \item \textbf{Compute standard focal loss:}
    \[
    \text{focal\_loss} = -\beta \cdot (1 - p_t)^{\gamma} \cdot \log(p_t + \epsilon)
    \]
    where 
    \[
    p_t = y_{\text{pred\_sigmoid}} \cdot y_{\text{true}} + (1 - y_{\text{pred\_sigmoid}}) \cdot (1 - y_{\text{true}})
    \]

    \item \textbf{Compute hard and easy losses:}
    \[
    \text{hard\_loss} = \sum (\text{focal\_loss} \times \text{hard}_{\text{dilated}})
    \]
    \[
    \text{easy\_loss} = \sum (\text{focal\_loss} \times \text{easy})
    \]

    \item \textbf{Calculate sample difficulty:}
    \[
    \text{sample\_difficulty} = 1.0 - \text{mean}(y_{\text{pred}})
    \]

    \item \textbf{Calculate annotation variability:}
    \[
    \text{annotation\_variability} = \text{mean}(y_{\text{std}})
    \]

    \item \textbf{Compute gamma factor:}
    \[
    \gamma = \text{sample\_difficulty} + \text{annotation\_variability}
    \]

    \item \textbf{Adjust hard and easy losses:}
    \[
    \text{weighted\_hard\_loss} = \gamma \cdot \text{hard\_loss}
    \]
    \[
    \text{weighted\_easy\_loss} = \frac{1}{\gamma} \cdot \text{easy\_loss}
    \]

    \item \textbf{Compute final loss:}
    \[
    \text{LOSS} = \frac{\text{weighted\_easy\_loss} + \text{weighted\_hard\_loss}}{N_{\text{pixels}}}
    \]

    \item \textbf{Return final loss:}
    \[
    \text{Return LOSS}
    \]
\end{enumerate}

\section*{Experimental Setup}
\subsection*{Dataset acquisition and preprocessing}
\label{subsec1}
\begin{figure}[ht]
    \centering
    \includegraphics[width=0.8\linewidth]{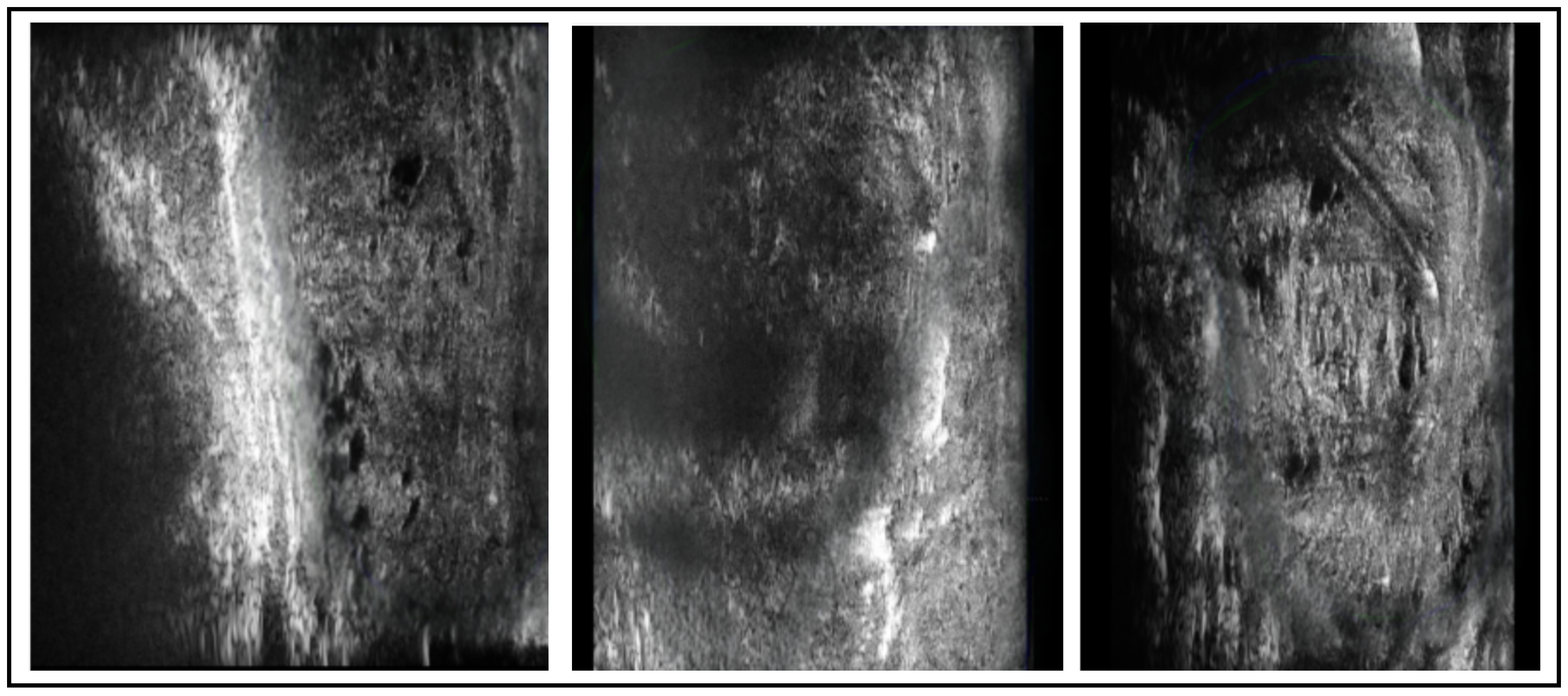}
    \caption{\textbf{Sample images of the dataset.}}
    \label{fig:Dataset_sample}
\end{figure}

This study utilizes the Micro-ultrasound Prostate Capsule Segmentation dataset\cite{jiang2024microsegnet}, which consists of micro-ultrasound images from 75 patients specifically curated for prostate segmentation tasks. The dataset is divided into a training set comprising 2,750 images from 55 patients and a testing set containing 600 images from 20 patients. Each patient's dataset includes approximately 200–300 micro-ultrasound images acquired in a pseudo sagittal plane. The images were normalized to an intensity range of [0, 1] and resized to 224x224 pixels to maintain a consistent scale and size for practical neural network training. Furthermore, various data augmentation techniques, such as random rotations, flips, and intensity variations, were applied to the training images. The dataset was annotated by experts and non-experts, with all annotations reviewed and refined to ensure high-quality ground truth. Annotation discrepancies were used to identify hard and easy regions incorporated into the AG-BCE loss function. Figure~\ref{fig:Dataset_sample} illustrates the sample images of the dataset.

\subsection*{Training setup}
The image patch size was set to 16, and a batch size of 8 was used for training. The learning rate was fixed at 0.01, with a momentum of 0.9 and a weight decay of $1 \times 10^{-4}$. These hyperparameters were selected through empirical testing and applied uniformly across all models. Each model was trained for a maximum of 10 epochs to reduce the risk of overfitting.

\subsection*{Evaluation metrics}

The DSC was employed to evaluate the overlap between the predicted prostate segmentation (\(p\)) and the ground truth segmentation (\(g\)) under a setup similar to the original experiment. It is defined as (equation 4):

\begin{equation}
\label{eq:DSC}
\text{DSC}(g, p) = \frac{2 \times |g \cap p|}{|g| + |p|}
\tag{4}
\end{equation}

Here, \(|g|\) and \(|p|\) denote the number of positive pixels in the ground truth and predicted segmentation, respectively, while \(g \cap p\) represents their intersection. The DSC ranges from 0 to 1, where higher values indicate greater overlap between the segmentations.

Additionally, the Hausdorff Distance (HD) was used to quantify the maximum distance between the boundaries of \(g\) and \(p\). The HD is given by (equation 5):

\begin{equation}
\text{HD} = \max \left\{ \sup_{G \in g} \inf_{P \in p} d(G, P), \sup_{P \in p} \inf_{G \in g} d(G, P) \right\}
\tag{5}
\end{equation}

To minimize the influence of small outliers, the 95th percentile of these distances (HD95) was adopted instead of the maximum distance.

\subsection*{Assumptions and validations}

The focal loss function addresses imbalanced datasets and challenging regions by dynamically adjusting the contribution of each example to the overall loss. This approach prioritizes harder examples while reducing the impact of easier ones, enhancing the model's generalization in complex scenarios. The computations for both the standard focal loss (\texttt{pyT\_focal\_loss}) and the adaptive focal loss assume the formula (equation 6):

\begin{equation}
\text{focal\_loss}(y_{\text{true}}, y_{\text{pred\_sigmoid}}) = -\beta \cdot (1 - p_t)^{\gamma} \cdot \log(p_t + \epsilon),
\tag{6}
\end{equation}

where \(p_t\) is the predicted probability for the true class, calculated as (equation 7):

\begin{equation}
p_t = y_{\text{pred\_sigmoid}} \cdot y_{\text{true}} + (1 - y_{\text{pred\_sigmoid}}) \cdot (1 - y_{\text{true}}),
\tag{7}
\end{equation}

The parameter \(\gamma\) controls the focus on hard versus easy regions, while \(\epsilon\) avoids errors due to logarithms of zero. This loss function helps the model focus on challenging cases, especially beneficial for imbalanced datasets or tasks with complex regions.

\subsubsection*{Bias considerations}

The focal loss functions aim to handle regions of varying difficulty with minimal bias. For harder regions, where predictions are uncertain or near decision boundaries, the \((1 - p_t)^{\gamma}\) term amplifies the loss, drawing the model’s attention. In contrast, the loss for easier regions is naturally smaller, enabling the model to concentrate less on well-classified examples. Adaptive focal loss additionally adjusts for annotation variability, dynamically modifying the \(\gamma\) factor to reflect the unique challenges of each sample.

\subsubsection*{Dataset characteristics}

The performance of these loss methods depends on dataset characteristics, such as class imbalance or inconsistent labels. The standard focal loss addresses class imbalance by emphasizing hard examples, while adaptive focal loss incorporates annotation variability to handle inconsistently labelled regions more effectively.

\subsubsection*{Validation of assumptions}

The assumptions underlying these focal loss methods were validated by comparing their performance against the AG-BCE loss function used in this study. The results confirm that these methods improve learning for the specific prostate segmentation task and dataset. Detailed comparisons are provided in the subsequent section.

\section*{Results and Discussion}

\subsection*{Loss Curve During Training}

Table~\ref{table:loss_comparison} provides a comparative analysis of loss values recorded over ten epochs for three distinct loss functions: standard focal loss (i.e., \texttt{pyT\_focal\_loss}), \texttt{adaptive\_focal\_loss}, and \texttt{AG\_BCE\_loss}. The findings indicate a consistent reduction in loss values across all three functions. Notably, the \texttt{adaptive\_focal\_loss} function achieves the lowest loss values, ranging from 0.212186 to 0.040419, underlining its superior capability in addressing complex regions and challenging cases. In contrast, the standard focal loss demonstrates a steady decline in loss values from 0.152941 to 0.128173, although its fixed focusing parameter results in slightly higher losses compared to the adaptive focal loss. Lastly, the \texttt{AG\_BCE\_loss} shows relatively higher initial loss values, decreasing from 0.106898 to 0.029263. These observations underscore the effectiveness of the adaptive focal loss's dynamic weighting mechanism in optimizing performance.

\begin{figure}[htbp]
    \centering
    \includegraphics[width=0.7\textwidth]{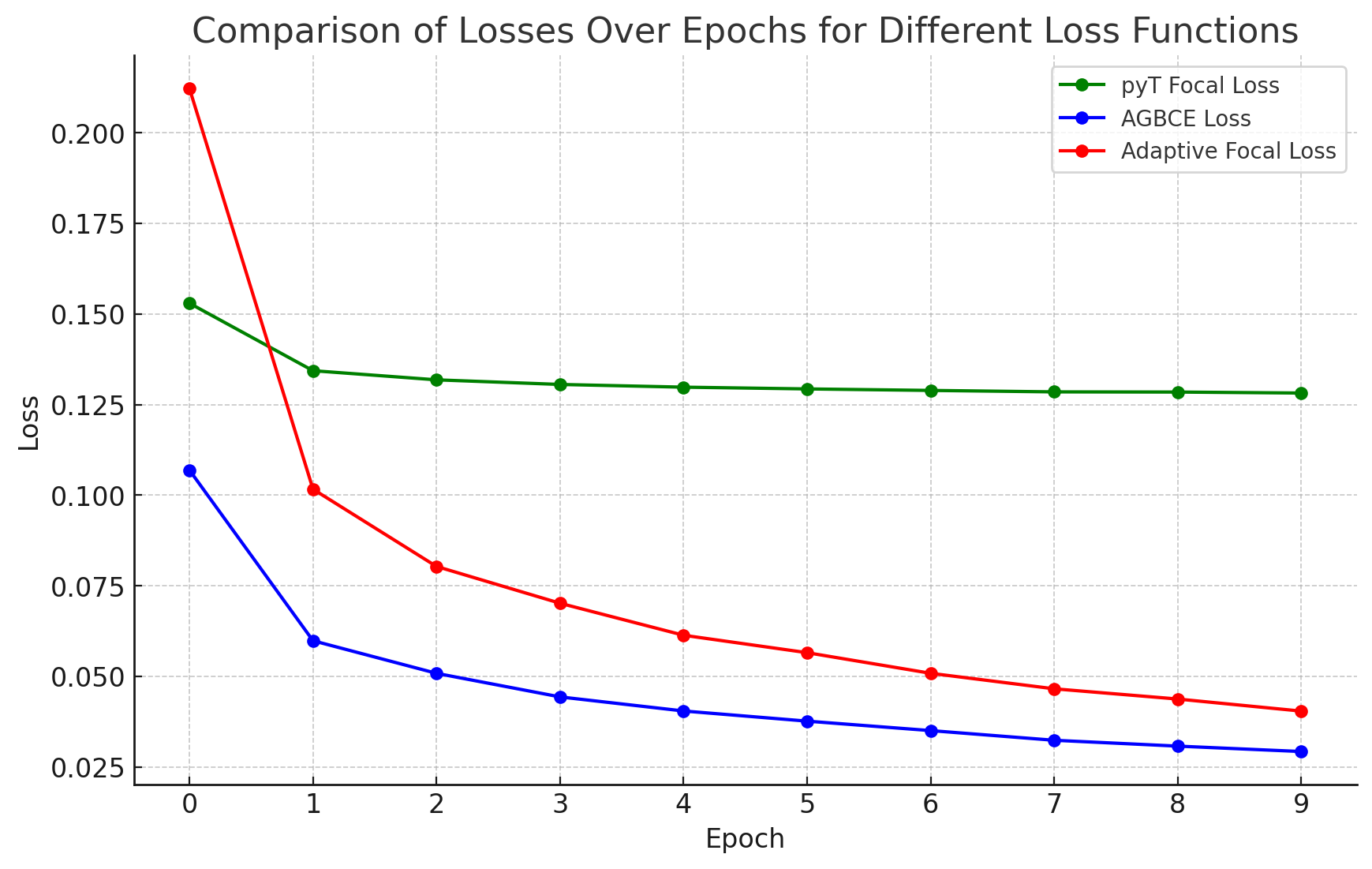}
    \caption{\textbf{Loss curve during training.} The graph illustrates the progression of loss values for all three loss functions across epochs, comparing their convergence rates and stability.}
    \label{fig:output_loss}
\end{figure}

Figure~\ref{fig:output_loss} illustrates the progression of the loss curve during training for all loss functions, highlighting differences in convergence rates and stability across epochs.

\begin{table}[htbp]
\centering
\caption{Comparison of loss values across epochs for different loss functions during training.}
\label{table:loss_comparison}
\small 
\renewcommand{\arraystretch}{1.2} 
\setlength{\tabcolsep}{4pt} 
\begin{tabular}{@{}cccc@{}}
\toprule
\textbf{Epoch} & \textbf{Adaptive Focal Loss} & \textbf{PyTorch Focal Loss} & \textbf{AG BCE Loss} \\ \midrule
1  & 0.212186 & 0.152941 & 0.106898 \\
2  & 0.101646 & 0.134336 & 0.059834 \\
3  & 0.080325 & 0.131814 & 0.050836 \\
4  & 0.070172 & 0.130542 & 0.044332 \\
5  & 0.061344 & 0.129793 & 0.040449 \\
6  & 0.056536 & 0.129307 & 0.037647 \\
7  & 0.050847 & 0.128894 & 0.035039 \\
8  & 0.046570 & 0.128493 & 0.032370 \\
9  & 0.043747 & 0.128443 & 0.030755 \\
10 & 0.040419 & 0.128173 & 0.029263 \\ \bottomrule
\end{tabular}
\end{table}

\noindent Table~\ref{table:comparison_performance} summarizes the Dice Similarity Coefficient (DSC) and Hausdorff Distance (HD) metrics obtained using three different loss functions, averaged across multiple test cases. Among the evaluated methods, the Adaptive Focal Loss Function demonstrated superior performance compared to the PyTorch\_Focal\_Loss and AG\_BCE\_Loss functions, excelling in both DSC and HD metrics. 

Specifically, the Adaptive Focal Loss Function improved the DSC from 0.913972 to 0.940013 when compared to the PyTorch Focal Loss Function, while reducing the HD from 2.497254~mm to 1.948677~mm. Although the numerical improvements may appear modest, they hold significant practical importance. The Adaptive Focal Loss Function achieves these enhancements by dynamically incorporating sample difficulty and annotation variability into the loss computation. This mechanism adjusts the contribution of each pixel to the loss based on prediction confidence and the unique challenges presented by each sample, ensuring more precise segmentation, particularly in regions with high complexity or variability. Such improvements are critical in medical imaging tasks, such as prostate segmentation, where even slight gains in boundary accuracy can substantially impact diagnostic reliability. The observed improvements in DSC and HD underscore the function's ability to effectively manage outliers and data variability, resulting in more accurate segmentations.

In contrast, while the AG-BCE Loss Function is designed to distinguish between hard and easy regions, it lacks the modulating factor that adjusts for prediction confidence. Despite this limitation, the AG-BCE Loss Function outperformed the PyTorch Focal Loss Function, achieving a DSC of 0.939764 and an HD of 2.000584~mm. However, it still fell short of the performance achieved by the Adaptive Focal Loss Function. These results highlight the importance of integrating adaptive mechanisms that respond to the specific characteristics of the data.

In conclusion, the Adaptive Focal Loss Function's ability to dynamically adjust based on prediction confidence and sample variability leads to more accurate and consistent segmentation outcomes. These findings suggest that further advancements in loss function design, incorporating similar adaptive principles, could pave the way for enhanced performance in challenging image segmentation applications.

\begin{table}[htbp]
\centering
\caption{\textbf{Comparison of performance metrics across 20 test cases for different loss functions.} The table presents the Mean Dice and Mean HD95 metrics for each test case and the average across all cases for three loss functions: Adaptive Focal Loss, pyTorch Focal Loss, and AG BCE Loss.}
\resizebox{\textwidth}{!}{%
\begin{tabular}{lcc|cc|cc}
\toprule
\textbf{Index} & \multicolumn{2}{c|}{\textbf{Adaptive Focal Loss}} & \multicolumn{2}{c|}{\textbf{pyTorch Focal Loss}} & \multicolumn{2}{c}{\textbf{AG BCE Loss}} \\
 & \textbf{Mean Dice} & \textbf{Mean HD95} & \textbf{Mean Dice} & \textbf{Mean HD95} & \textbf{Mean Dice} & \textbf{Mean HD95} \\
\midrule
1  & 0.9569 & 1.9925 & 0.9317 & 3.0807 & 0.9525 & 2.6741 \\
2  & 0.9254 & 2.3482 & 0.9396 & 2.1233 & 0.9118 & 3.1074 \\
3  & 0.9576 & 1.1166 & 0.9160 & 1.7192 & 0.9499 & 1.1531 \\
4  & 0.9421 & 1.8843 & 0.9046 & 2.8805 & 0.9417 & 1.9182 \\
5  & 0.9381 & 1.5865 & 0.9133 & 2.2564 & 0.9336 & 1.7901 \\
6  & 0.9344 & 2.6164 & 0.9004 & 3.1953 & 0.9395 & 2.4522 \\
7  & 0.9554 & 1.4290 & 0.9408 & 1.7067 & 0.9606 & 1.2860 \\
8  & 0.9577 & 1.2766 & 0.9284 & 1.9655 & 0.9613 & 1.1483 \\
9  & 0.8939 & 2.6147 & 0.8445 & 3.1875 & 0.8525 & 3.3892 \\
10 & 0.9550 & 0.9854 & 0.9166 & 1.6644 & 0.9479 & 1.1115 \\
11 & 0.9426 & 2.1823 & 0.9194 & 2.9209 & 0.9436 & 2.3633 \\
12 & 0.9150 & 2.4519 & 0.8860 & 2.9951 & 0.9244 & 2.0885 \\
13 & 0.9244 & 2.0687 & 0.8987 & 2.4553 & 0.9374 & 1.7690 \\
14 & 0.9403 & 2.1641 & 0.8905 & 3.0592 & 0.9458 & 2.2211 \\
15 & 0.9539 & 1.5215 & 0.9435 & 1.7696 & 0.9569 & 1.4332 \\
16 & 0.9341 & 2.1774 & 0.9050 & 2.5957 & 0.9457 & 1.9767 \\
17 & 0.9467 & 2.4147 & 0.9168 & 3.9691 & 0.9488 & 2.2691 \\
18 & 0.9547 & 1.6909 & 0.9299 & 2.2262 & 0.9541 & 1.6121 \\
19 & 0.9542 & 1.4059 & 0.9134 & 1.8777 & 0.9510 & 1.3725 \\
20 & 0.9178 & 3.0461 & 0.9403 & 2.2969 & 0.9364 & 2.8758 \\
\midrule
\textbf{Mean} & \textbf{0.9400} & \textbf{1.9487} & \textbf{0.9140} & \textbf{2.4973} & \textbf{0.9398} & \textbf{2.0006} \\
\bottomrule
\end{tabular}
}
\label{table:comparison_performance}
\end{table}

\subsection*{Training parameters and model performance}

\begin{figure}[ht]
    \centering
    \vspace{-0.5cm} 
    \includegraphics[width=0.8\textwidth]{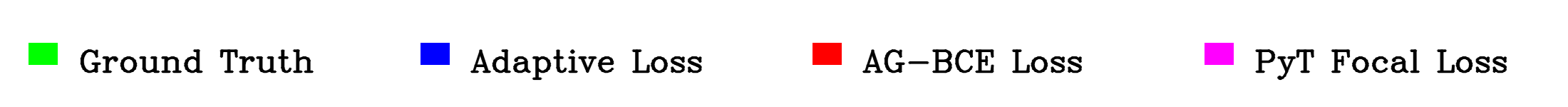} \\
    \vspace{0.3cm} 
    \begin{tabular}{ccc} 
        \textbf{Test Image 1} & \textbf{Test Image 2} & \textbf{Test Image 3} \\
        
        \includegraphics[width=0.28\textwidth]{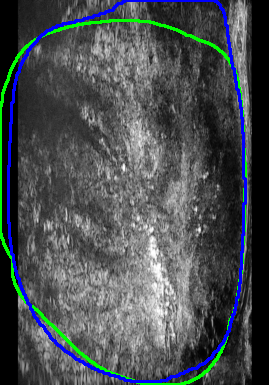} &
        \includegraphics[width=0.28\textwidth]{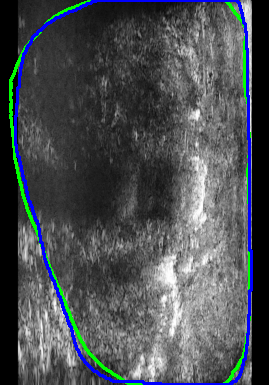} &
        \includegraphics[width=0.28\textwidth]{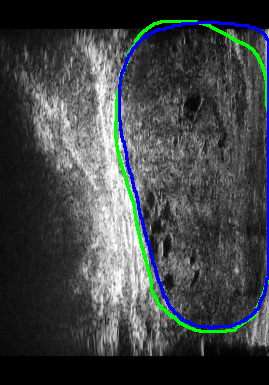} \\
        
        \includegraphics[width=0.28\textwidth]{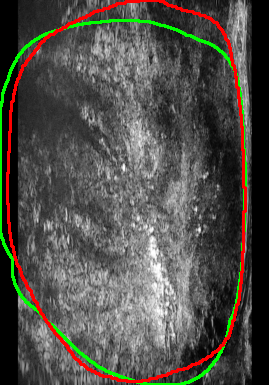} &
        \includegraphics[width=0.28\textwidth]{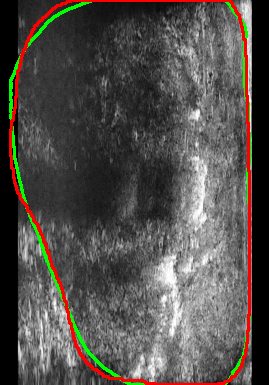} &
        \includegraphics[width=0.28\textwidth]{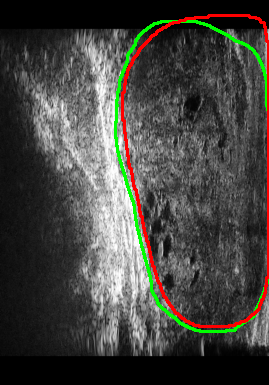} \\
        
        \includegraphics[width=0.28\textwidth]{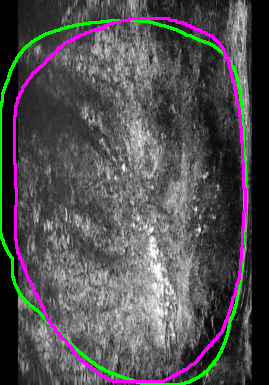} &
        \includegraphics[width=0.28\textwidth]{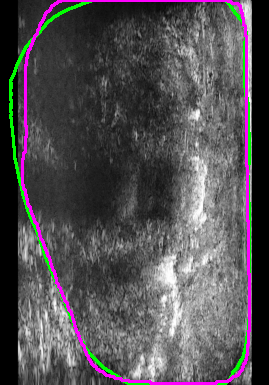} &
        \includegraphics[width=0.28\textwidth]{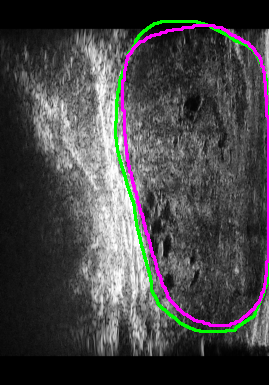} \\
    \end{tabular}
    \vspace{0.3cm} 
    \caption{\textbf{Segmentation comparison of different loss functions.} Segmentation results on three test images using Adaptive Focal Loss, AG-BCE Loss, and PyTorch Focal Loss. Each row shows the outputs for one loss function, and the columns correspond to different test images. The legend at the top indicates the colour coding for ground truth and segmentation outputs.}
    \label{fig:segmentation_comparison}
\end{figure}

Figure~\ref{fig:segmentation_comparison} presents the performance evaluation of three loss functions—Adaptive Focal Loss, AG-BCE Loss, and PyTorch Focal Loss—on three test images. Each row in the figure corresponds to the segmentation results of one loss function, while the columns represent the respective test images. A legend at the top provides the color coding for the ground truth and the outputs of the three loss functions, allowing for a direct visual comparison of their performance.

\subsection*{Novelty of the Adaptive Function}

Traditional loss functions, such as PyTorch's Focal Loss, assume equal contributions from all image regions. However, this assumption falls short in medical imaging, where critical areas are harder to segment due to noise, artifacts, or subtle boundaries. The AG-BCE loss function attempts to address this by assigning greater weight to complex regions; however, it does so in a static manner, applying the same weight adjustments across all samples. This static approach fails to capture the dynamic nature of difficulty within and between images.

In contrast, the Adaptive Focal Loss function addresses these challenges by modifying the loss contribution based on the model's forecast confidence, dynamically adjusting the focus to accommodate the unique difficulties of each image or region. This adaptability is achieved by introducing a focusing parameter ($\gamma$), which emphasizes hard-to-classify data, and a balancing component ($\alpha$), which mitigates class imbalance. Adaptive Focal Loss enhances model performance by focusing on the most challenging and clinically relevant areas.

Segmenting prostate borders in micro-ultrasound (micro-US) imaging presents numerous unique challenges. Although micro-US offers superior resolution compared to conventional ultrasonography, it often suffers from severe speckle noise and low contrast, particularly in regions where the prostate boundary is less defined. Variability in tissue density and calcifications further complicate segmentation, as these factors may introduce false signals that confuse the model. Additionally, common artifacts arising from the interaction of sound waves with varying tissue types and densities can distort the true boundaries of the prostate, leading to inaccurate segmentation. Prostate deformation caused by patient movement or pressureffectively handle regions with low contrast, excessive noise, or significant anatomical variations images, hindering the model's ability to generalize.

To address these challenges, the Adaptive Focal Loss function dynamically adjusts the learning focus based on the unique obstacles posed by each sample. By varying the contribution of each pixel to the loss based on prediction confidence, the model is encouraged to allocate more learning capacity to challenging cases where segmentation is most complex. This approach enables the model to handle regions with low contrast, excessive noise, or significant anatomical variation more effectively. Moreover, the dynamic adjustment of the loss function through the focusing parameter allows the model to tackle artifacts and distortions in micro-US images efficiently. Unlike static methods, this approach recognizes that the importance of different data regions varies depending on the specific conditions of each image. This context-aware strategy enables the model to better interpret the data, thereby improving generalization to unseen cases.

\begin{figure}[htbp]
    \centering
    \includegraphics[width=0.8\textwidth]{legend.png}
    \vspace{0.3cm}
    \begin{tabular}{ccc}
        \textbf{Test Image 04} & \textbf{Test Image 11} & \textbf{Test Image 14} \\
        \includegraphics[width=0.28\textwidth]{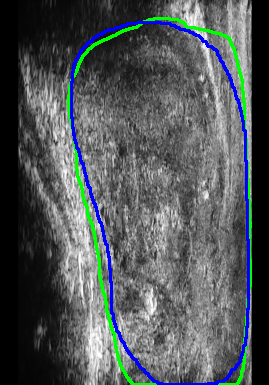} &
        \includegraphics[width=0.28\textwidth]{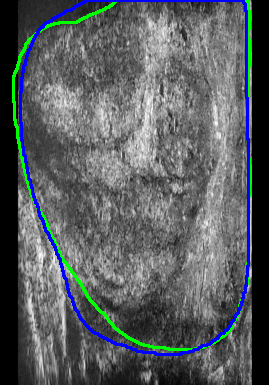} &
        \includegraphics[width=0.28\textwidth]{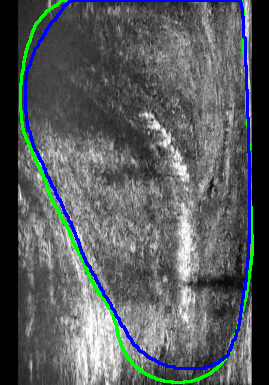} \\
        \includegraphics[width=0.28\textwidth]{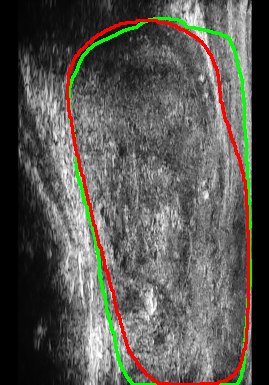} &
        \includegraphics[width=0.28\textwidth]{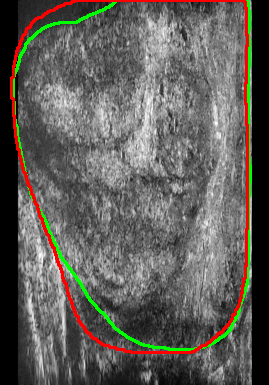} &
        \includegraphics[width=0.28\textwidth]{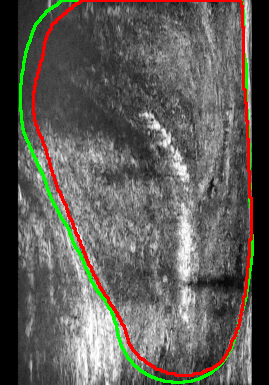} \\
        \includegraphics[width=0.28\textwidth]{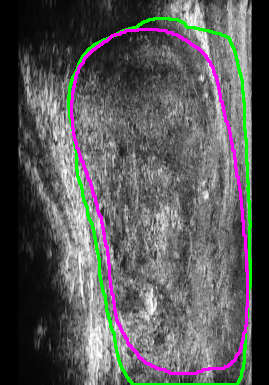} &
        \includegraphics[width=0.28\textwidth]{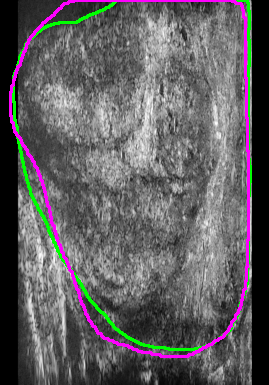} &
        \includegraphics[width=0.28\textwidth]{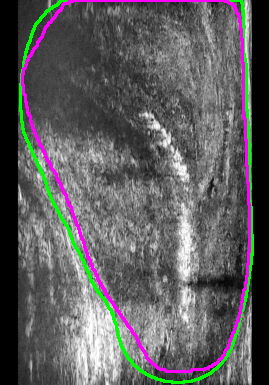} \\
    \end{tabular}
    \caption{\textbf{Comparison of segmentation results on test images 04, 11, and 14.} Segmentation results are presented for Adaptive Focal Loss, AG-BCE, and pyT Focal Loss across three test images. Each row corresponds to a different loss function, as indicated in the legend.}
    \label{fig:segmentation_comparison_1}
\end{figure}

The visual analysis of segmentation results, illustrated in Figure~\ref{fig:segmentation_comparison_1}, demonstrates that Adaptive Focal Loss consistently achieves more accurate segmentation, particularly in regions with low contrast, high noise, or complex anatomical variations, closely aligning with the ground truth.

\begin{enumerate}
    \item \textbf{Test Image 04}: The Adaptive Focal Loss function excelled in capturing the prostate boundary, especially in the lower-left region obscured by speckle noise. By comparison, AG-BCE and PyTorch Focal Loss exhibited significant deviations, notably under-segmenting the prostate and failing to depict the boundary accurately.

    \item \textbf{Test Image 11}: The Adaptive Focal Loss function demonstrated superior performance in the upper-right region, where the prostate boundary was subtly defined against surrounding tissue. AG-BCE failed to capture finer details, producing a less smooth boundary and shifting it inward, while PyTorch Focal Loss over-segmented, extending beyond the prostate.

    \item \textbf{Test Image 14}: Despite complex artifacts in the lower region, Adaptive Focal Loss maintained a boundary closely resembling the ground truth. AG-BCE exhibited minor mismatches, particularly in the lower-right corner, while PyTorch Focal Loss underperformed, displaying over-segmentation and boundary deviations.
\end{enumerate}

\begin{figure}[htbp]
    \centering
    \includegraphics[width=0.8\textwidth]{legend.png}
    \vspace{0.3cm}
    \begin{tabular}{ccc}
        \textbf{Test Image 17} & \textbf{Test Image 18} & \textbf{Test Image 19} \\
        \includegraphics[width=0.28\textwidth]{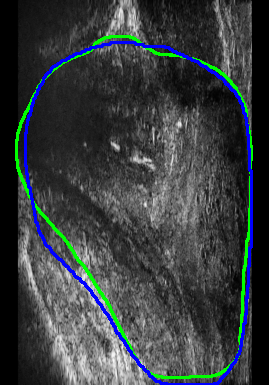} &
        \includegraphics[width=0.28\textwidth]{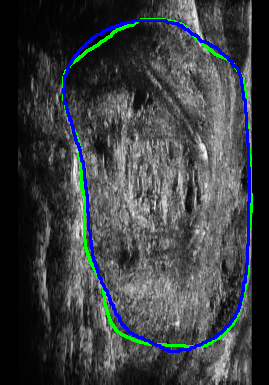} &
        \includegraphics[width=0.28\textwidth]{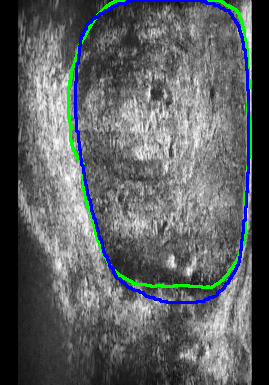} \\
        \includegraphics[width=0.28\textwidth]{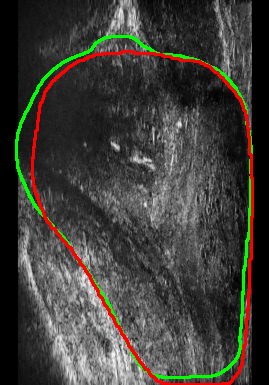} &
        \includegraphics[width=0.28\textwidth]{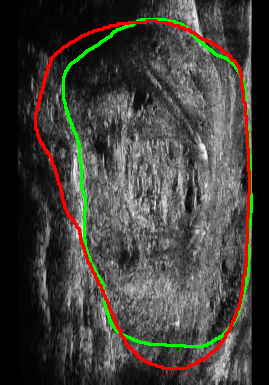} &
        \includegraphics[width=0.28\textwidth]{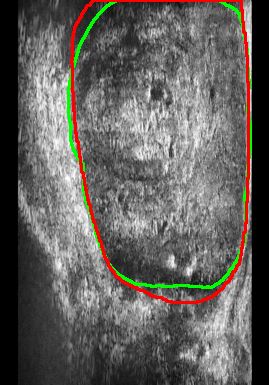} \\
        \includegraphics[width=0.28\textwidth]{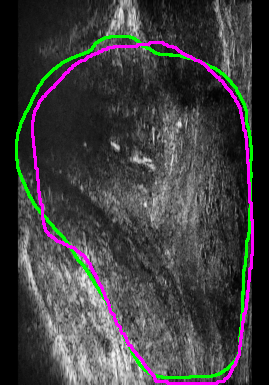} &
        \includegraphics[width=0.28\textwidth]{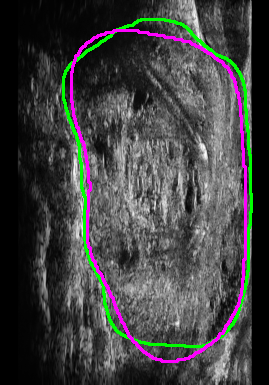} &
        \includegraphics[width=0.28\textwidth]{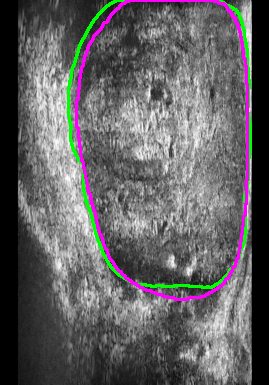} \\
    \end{tabular}
    \caption{\textbf{Comparison of segmentation results on test images 17, 18, and 19.} Segmentation results are presented for Adaptive Focal Loss, AG-BCE, and pyT Focal Loss across three test images. Each row corresponds to a different loss function, as indicated in the legend.}
    \label{fig:segmentation_comparison_2}
\end{figure}

Further comparisons of Test 17, Test 18, and Test 19, shown in Figure~\ref{fig:segmentation_comparison_2}, underscore the effectiveness of the Adaptive Focal Loss in handling challenging prostate boundary regions.

\begin{enumerate}
    \item \textbf{Test Image 17:} The Adaptive Focal Loss performed exceptionally well in the upper left region, which presented significant anatomical deformation, making segmentation particularly difficult. In contrast, AG-BCE produced an overly smooth boundary, while PyTorch Focal Loss severely under-segmented the area.
    \item \textbf{Test Image 18:} This image highlighted challenges in the lower prostate boundary caused by artifacts and low contrast. The Adaptive Focal Loss accurately delineated the boundary, whereas AG-BCE exhibited boundary shifts and PyTorch Focal Loss resulted in under-segmentation.
    \item \textbf{Test Image 19:} The Adaptive Focal Loss again outperformed the other methods, especially in the upper right region, where tissue density variations blurred the boundary. AG-BCE over-segmented this region, and PyTorch Focal Loss produced inconsistent results, with instances of both under- and over-segmentation.
\end{enumerate}

Overall, the Adaptive Focal Loss function consistently demonstrated superior performance in accurately delineating prostate boundaries in challenging micro-ultrasound segmentation tasks, outperforming both AG-BCE and PyTorch Focal Loss in accuracy.

\section*{Limitations and Future Work}
The Adaptive Focal Loss function has improved segmentation accuracy, particularly in challenging regions; however, it has some limitations. One significant issue is the increased computational complexity arising from adjusting weights based on sample difficulty and annotation variability. This additional computational cost may not be suitable for environments with limited resources. Moreover, in datasets with low noise or variability, the adaptive method might not provide substantial advantages over simpler loss functions that are easier to implement and computationally faster. 

Another limitation is the reliance on accurately estimating sample difficulty. Errors in this estimation process can adversely affect the model's performance, unlike methods with fixed weights that are less sensitive to such errors. Furthermore, while the Adaptive Focal Loss has demonstrated promising results for micro-ultrasound images, its effectiveness in other medical imaging modalities remains uncertain. In such cases, alternative loss functions may be more appropriate.

Future work will enhance the adaptive mechanism to mitigate over-sensitivity to extreme cases of sample difficulty and variability. This could involve developing a more refined weighting system to distinguish significant variability from outlier noise. A potential avenue for exploration is the integration of meta-learning techniques to adaptively adjust the loss function based on the dataset's characteristics. Additionally, evaluating Adaptive Focal Loss in other medical imaging fields, such as MRI and CT scans, will help establish its generalizability across various imaging modalities.

Another key research direction involves testing the concept of dilating hard regions to enable the model to learn more accurate boundaries in complex areas. This approach could improve segmentation performance in regions with intricate shapes. However, as this concept was not directly tested in the current study, further investigation is necessary to determine whether dilation enhances segmentation accuracy or introduces complications in regions with clearer boundaries. Addressing these limitations and exploring these future directions will enable the Adaptive Focal Loss to be applied more effectively across a broader range of medical imaging tasks.

\section*{Conclusion}
This paper discusses the significant improvements achieved by applying the Adaptive Focal Loss Function for prostate capsule segmentation using micro-ultrasound images. The Adaptive Focal Loss Function demonstrates exceptional performance, with loss values decreasing from 0.212186 to 0.040419 over ten epochs, highlighting its effectiveness in handling complex and challenging regions. 

The study uses a micro-ultrasound dataset, which is susceptible to noise and variability, making the adaptive approach essential for accurate segmentation. The dataset, created explicitly for prostate segmentation, includes images from 75 patients, with 2,750 training images from 55 patients and 600 testing images from 20 patients. Each patient's dataset contains 200–300 micro-ultrasound images in a pseudo-sagittal plane. To ensure consistency in scale and size, the images were normalized to an intensity range of [0, 1] and resized to 224x224 pixels for practical neural network training.

Data augmentation techniques such as random rotations, flips, and intensity variations were also applied to the training images. Annotations were provided by both experts and non-experts, with all annotations reviewed and refined to ensure high-quality ground truth. Discrepancies in annotations were used to identify challenging and simple regions incorporated into the AG-BCE loss function. 
The results demonstrate that the Adaptive Focal Loss consistently outperforms traditional loss functions, such as AG-BCE and PyTorch Focal Loss, particularly in regions with low contrast and high noise. This context-aware method improves data interpretation, leading to more accurate segmentation results that align closely with ground truth. 
Overall, the findings highlight the potential of the Adaptive Focal Loss Function to improve medical imaging tasks, especially in complex segmentation scenarios.

\section*{CRediT Author Contribution Statement}
\textbf{Kaniz Fatema:} Conceptualization, Methodology, Writing, Review \& Editing. \textbf{Vaibhav Thakur:} Methodology, Software, Writing. \textbf{Emad A. Mohammed:} Supervision, Methodology, Validation, Review \& Editing.

\section*{Declaration of Competing Interests}
The authors declare no competing financial interests or personal relationships that could influence the work reported in this paper.

\section*{Data Availability}
The data utilized in this paper is publicly available from \cite{jiang2024microsegnet}. The code supporting this paper is available at the following GitHub repository \cite{adaptiveFocalLoss}.

\end{document}